# SECURING A MOBILE ADHOC NETWORK FROM ROUTING ATTACKS THROUGH THE APPLICATION OF GENETIC ALGORITHM

Kumar Nikhil[1] and Swati Agarwal[2] and Pankaj Sharma[3]
[1]Department of IT, *ABES Engineering College, Ghaziabad (U.P.), India*
kumar_nikhill20@rediff.com
[2]Department of IT, *ABES Engineering College, Ghaziabad (U.P.), India*
swatiabes08@gmail.com
[3]Sr. Asstt. Professor (IT), *ABES Engineering College, Ghaziabad (U.P.), India*
sharma1.pk@gmaill.com

***Abstract:*** *In recent years, the static shortest path (SP) problem has been well addressed using intelligent optimization techniques, e.g., artificial neural networks, genetic algorithms (GAs), particle swarm optimization, etc. However, with the advancement in wireless communications, more and more mobile wireless networks appear, e.g., mobile networks [mobile ad hoc networks (MANETs)], wireless sensor networks, etc. One of the most important characteristics in mobile wireless networks is the topology dynamics, i.e., the network topology changes over time due to energy conservation or node mobility. Therefore, the SP routing problem in MANETs turns out to be a dynamic optimization problem. GA's are able to find, if not the shortest, at least an optimal path between source and destination in mobile ad-hoc network nodes. And we obtain the alternative path or backup path to avoid reroute discovery in the case of link failure or node failure.*

## 1. INTRODUCTION

A mobile *ad hoc* network *(MANET)* is an autonomous network that consists of mobile nodes that communicate with each other over wireless links. This type of networks is suited for use in situations where a fixed infrastructure is not available, not trusted, too expensive or unreliable. A few examples include: a network of notebook computers or *PDAs* in a conference or campus setting, rescue operations, and headquarters industry. In the absence of a fixed infrastructure, nodes have to cooperate in order to provide the necessary network functionality. Routing is one of the primary functions each node has to perform in order to enable connections between nodes that are not directly within each other's send range. The development of efficient routing protocols is a non trivial and challenging task because of the specific characteristics of a MANET environment:

• Due to node movements, the network topology may change randomly and rapidly at unpredicted times.

• The available bandwidth is limited and can vary due to fading, noise, interference.

• Most mobile devices are battery powered; therefore energy consumption plays an important role.

In *ad–hoc* networks nodes geographically close to each other are grouped into non overlapping sub networks, clusters. Each cluster has a leading node called the clusterhead and a number of cluster members. When a cluster member wants to communicate with another node, a route is provided by its clusterhead. A crucial question is which node will become a clusterhead. Typically a clusterhead is more burdened than its members and could easily become a bottleneck of the system if not chosen appropriately.

Hence solutions to this problem are based on heuristics approaches. A good clustering scheme should preserve its structure as much as possible, when nodes are moving and/or the topology is slowly changing. Otherwise, recompilation of cluster heads and frequent information exchange among the participating nodes will result in high computation overhead. Any node can become a clusterhead if it has the necessary functionality, such as processing and transmission power. Nodes register with the nearest clusterhead and become members of that cluster. Clusters may change dynamically, reflecting the mobility of the underlying network.

The rest of the paper is organized as follows: section 2 deals with routing in ad hoc networks. Section 3 introduces the genetic algorithm as an optimization technique. Section 4 includes the steps that are required to apply the Genetic algorithm. Section 5 contains routing optimization using genetic algorithm. Conclusions are summarized in section 6.

## 2. ROUTING ATTACKS IN AD-HOC NETWORKS

The malicious node(s) can attacks in MANET using different ways, such as sending fake messages several times, fake routing information, and advertising fake links to disrupt routing operations. In the following subsection, current routing attacks and its countermeasures against MANET protocols are discussed in detail.

**Flooding attack:** In flooding attack, attacker exhausts the network resources, such as bandwidth and to consume a node's resources, such as computational and battery power or to disrupt the routing operation to cause severe degradation in network performance. For example, in AODV protocol, a malicious node can send a large number of RREQs in a short period to a destination node that does not exist in the network. Because no one will reply to the RREQs, these RREQs will flood the whole network. As a result, all of the node battery power, as well as network bandwidth will be consumed and could lead to denial-of-service.
A simple mechanism proposed to prevent the flooding attack in the AODV protocol. In this approach, each node monitors and calculates the rate of its neighbors' RREQ. If the RREQ rate of any neighbor exceeds the predefined threshold, the node records the ID of this neighbor in a blacklist. Then, the node drops any future RREQs from nodes that are listed in the blacklist. The limitation of this approach is that it cannot prevent against the flooding attack in which the flooding rate is below the threshold. Another drawback of this approach is that if a malicious node impersonates the ID of a legitimate node and broadcasts a large number of RREQs, other nodes might put the ID of this legitimate node on the blacklist by mistake.

**Blackhole attack:** In a blackhole attack, a malicious node sends fake routing information, claiming that it has an optimum route and causes other good nodes to route data packets through the malicious one.
For example, in AODV, the attacker can send a fake RREP (including a fake destination sequence number that is fabricated to be equal or higher than the one contained in the RREQ) to the source node, claiming that it has a sufficiently fresh route to the destination node. This causes the source node to select the route that passes through the attacker. Therefore, all traffic will be routed through the attacker, and therefore, the attacker can misuse or discard the traffic. Figure 1 shows an example of a blackhole attack, where attacker A sends a fake RREP to the source node S, claiming that it has a sufficiently fresher route than other nodes. Since the attacker's advertised sequence number is higher than other nodes' sequence numbers, the source node S will choose the route that passes through node A.

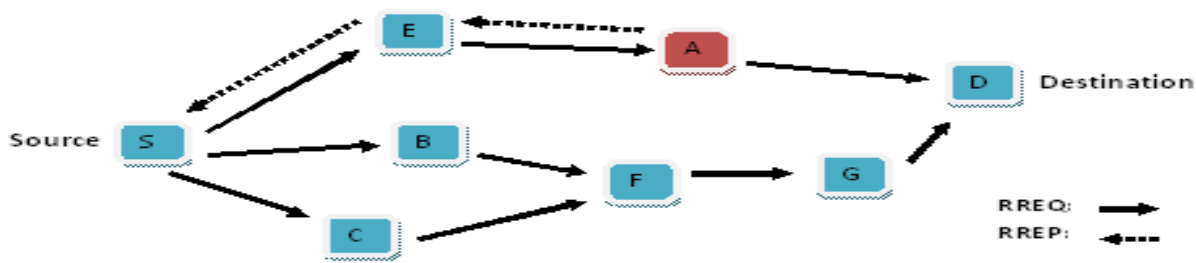

The route confirmation request (CREQ) and route confirmation reply (CREP) is introduced to avoid the blackhole attack. In this approach, the intermediate node not only sends RREPs to the source node but also sends CREQs to its next-hop node toward the destination node. After receiving a CREQ, the next-hop node looks up its cache for a route to the destination. If it has the route, it sends the CREP to the source. Upon receiving the CREP, the source node can confirm the validity of the path by comparing the path in RREP and the one in CREP. If both are matched, the source node judges that the route is correct. One drawback of this approach is that it cannot avoid the blackhole attack in which two consecutive nodes work in collusion, that is, when the next-hop node is a colluding attacker sending CREPs that support the incorrect path. In several reaearches, the authors proposed a solution that requires a source node to wait until a RREP packet arrives from more than two nodes. Upon receiving multiple RREPs, the source node checks whether there is a shared hop or not. If there is, the source node judges that the route is safe. The main drawback of this solution is that it introduces time delay, because it must wait until multiple RREPs arrive. In another attempt, the authors analyzed the blackhole attack and showed that a malicious node must increase the destination sequence number sufficiently to convince the source node that the route provided is sufficiently enough.

**Colluding misrelay attack:** In colluding misrelay attack, multiple attackers work in collusion to modify or drop routing packets to disrupt routing operation in a MANET. This attack is difficult to detect by using the conventional methods such as *watchdog* and *pathrater*. Figure 4 shows an example of this attack.

Consider the case where node A1 forwards routing packets for node T. In the figure, the first attacker A1 forwards routing packets as usual to avoid being detected by node T. However, the second attacker A2 drops or modifies these routing packets.

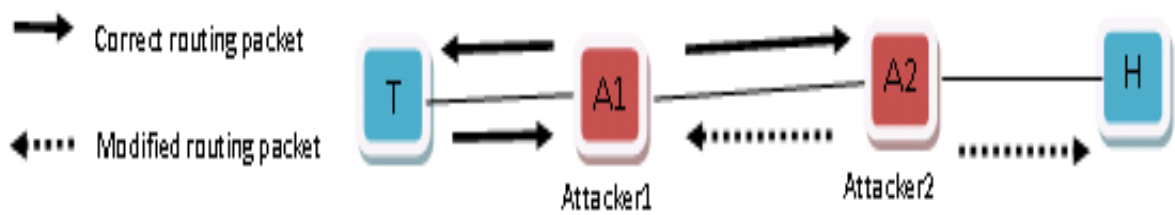

A conventional acknowledgment-based approach might detect this type of attack in a MANET, especially in a proactive MANET, but because routing packets destined to all nodes in the network require all nodes to return an ACK, this could lead to a large overhead, which is considered to be inefficient. In general, the main drawback of this approach is that even if we require each node to increase transmission power to be K times, we still cannot detect the attack in which K + 1 attackers work in collusion to drop packets.
Therefore, further work must be done to counter against this type of attack efficiently.

**Link spoofing attack:** In a link spoofing attack, a malicious node advertises fake links with non-neighbors to disrupt routing operations. For example, in the OLSR protocol, an attacker can advertise a fake link with a target's two-hop neighbors. Figure 2 shows an example of the link spoofing attack in an OLSR MANET. In the figure, we assume that node A is the attacking node, and node T is the target to be attacked. Before the attack, both nodes A and E are MPRs for node T. During the link spoofing attack, node A advertises a fake link with node T's two-hop neighbor, that is, node D. According to the OLSR protocol, node T will select the malicious node A as its only MPR since node A is the minimum set that reaches node T's two-hop neighbors. By being node T's only MPR, node A can then drop or withhold the routing traffic generated by node T.

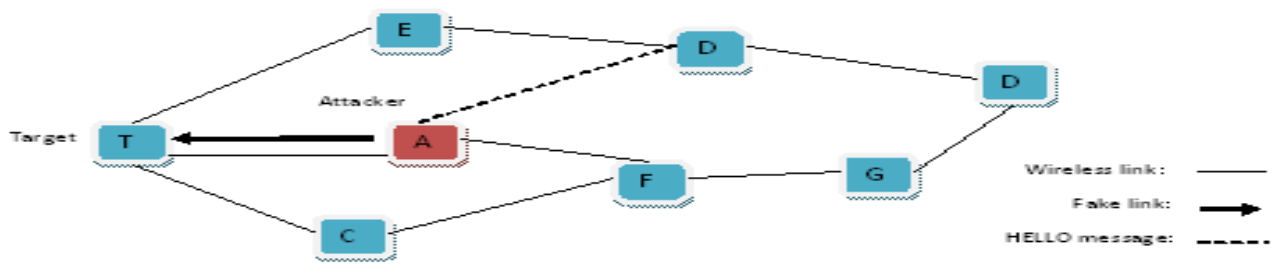

A location information-based detection method is proposed to detect link spoofing attack by using cryptography with a GPS and a time stamp. This approach requires each node to advertise its position obtained by the GPS and the time stamp to enable each node to obtain the location information of the other nodes. This approach detects the link spoofing by calculating the distance between two nodes that claim to be neighbors and checking the likelihood that the link is based on a maximum transmission range. The main drawback of this approach is that it might not work in a situation where all MANET nodes are not equipped with a GPS. Furthermore, attackers can still advertise false information and make it hard for other nodes to detect the attack.

## 3. GENETIC ALGORITHMS

The GA, which was introduced by John Holland, was adopted from natural evolution. Natural evolution has the following features:
1) The characteristics of an individual are encoded on a chromosome.
2) Each chromosome has certain fitness according to the environment in which it exists.
3) Individuals judged stronger are able to survive and produce next generations of strong individuals.

The GA is based on the above features in the following manner: the solution of the problem is encoded on a string comparable with the chromosome of the biological system.

The GA keeps a population of randomly selected chromosomes and allows filter chromosomes to combine and produce offspring with new characteristics, which may replace low fitness old chromosomes. This is repeated until we find a chromosome with best characteristics, which represents the optimal solution of the problem.

There are two mechanisms that link a genetic algorithm to the problem it is solving. These two mechanisms are:
1) Encoding solutions to the problem on chromosomes.

2) Evaluation function that returns a measurement of the worth of a chromosome in the context of the problem.

This is what we call the fitness of a chromosome. The evaluation function plays the same role in the genetic algorithm that the environment plays in natural evolution.

In order to use GA's for network topological design, the chromosome is chosen to contain the network parameters. A possible chromosome would be a string containing the weights of all nodes of the network. The evaluation function which assigns fitness to each chromosome is chosen according to the objective of the design problem. If the objective is to minimize the route between source and destination, then the evaluation function will compute the all distances of all possible paths between source and destination and give the dynamic optimal path with time change.

## 4. ROUTING OPTIMIZATION USING GENETIC ALGORITHM

The goal consists of allocate near optimal path from source to destination based on time, giving priority to cluster heads to maximize utilization and minimum delay.[5]

### Step1: Encoding and Initial Population

All nodes in the search space should be present and have a unique representation. If there is a one-to-one correspondence between the search space and string representation, the design of the genetic operator would be considerably less complex. These unique id's are used to encode the chromosome using integer permutation Encoding the individual chromosomes is an essential part of the mapping process; each chromosome contains information about the clusterheads and the members thereof, as obtained from the original clustering algorithm.

Each chromosome is represented by a link weight vector $W = <w1....w(n)>$ where *(n)* is the total number of links in the network. The value of each weight is within the range from *1* to *MAX_WEIGHT*. We define the value of *MAX_WEIGHT* to be *64* for reducing the search space. The population size is set to 100, with the initial values inside each chromosome randomly varying from *1* to *MAX_WEIGHT*.

### Step2: Fitness Evaluation

Chromosomes are selected according to their fitness. The bandwidth constraint is embedded into the fitness function as a penalty factor, such that the search space is explored with potential feasible solution. The fitness of each chromosome can be defined to be a two-dimensional function as shown in Equation 1. The overall network load *(L1)* and excessive bandwidth allocated to overloaded links

*(L2).*

Fitness = f(A1,A2) =c/(a x L1 + bxL2)        (1)

Where *a , b* and c are manually configured coefficients.
*A1* and *A2* are expressed as shown in equations 2, 3, 4.

$A1 = \sum_{g=1}^{G} D_g$        (2)
$A2 = \sum_{(I,j)\epsilon\sum} W_{ij} \times (\sum_{g=1}^{G} D_g - C_{ij})$        (3)
Where
$W_{ij} = \{0 \text{ if } \sum^{G} D_g <= C_{ij})$
   or {1 otherwise        (4)

Where:
*Dg:* Bandwidth demand for *cluster g* on each link;
*Cij:* Bandwidth capacity of link *(i,j)*;
*G:* total number of active cluster*s.*
So the objective function is two fields: first chromosomes of the new generations. And second, solutions obtained from the offspring should be feasible in that the total bandwidth allocated flows traveling through each link should not exceed its capacity. The tuning of ά and β can be regarded as a tradeoff between overall bandwidth conservation and load balancing. For example we let β = 0 then the objective is to conserve bandwidth resources only, while setting ά = 0 infers to minimize link overloading within the network.

### Step3: Crossover and Mutation
According to the basic principle of Genetic algorithms**,** chromosomes with better fitness value have higher probability of being inherited into the next generation. To achieve this, first we rank all the chromosomes in descending order to their fitness, so the chromosomes with high fitness (lower overall load) are placed on the top of the ranking list. Then we partition this list into two disjointed sets, with the top 50 chromosomes belonging to the upper class (*UC*) and the bottom 50 chromosomes to the lower class (*LC*). During the crossover procedure, we select one parent chromosome $C_u^i$ from UC and other parent $C_i^i$ from *LC* in generation "*i*" for creating the child $C^{i+1}$ in generation *i+1*. We use a crossover probability threshold $K_c \ \varepsilon \ (0.05)$ to decide the genes of which parent to be inherited into the child chromosome in the next generation. We introduce a mutation probability threshold $K_M$ to randomly replace some old genes with new ones.

## 5. CONCLUSIONS
We presented a genetic algorithm as an optimization technique for minimizing attacks in *MANET*. The results show that, with the genetic algorithmic technique each clusterhead handles the maximum possible number of mobile nodes in its cluster in order to facilitate the optimal operation of the medium access control (*MAC*) protocol, reduce the number of clusters and hence cluster heads, as well as, the loads among clusters are more evenly balanced by factor of ten.
A genetic algorithm technique mapped the possible solutions given by a weight based distributed clustering algorithm in order to find the better solution from a pool of solutions. Each clusterhead handles the maximum possible number of nodes in its cluster. Also a fewer cluster heads are obtained by the genetic algorithm technique. With the genetic algorithm technique the cumulative distributions of the paths are almost the same. Generally, another criterion of research can concentrate to simplify parameters of GA's optimization to leave the bad one out and optimize the good parameters. The genetic algorithm can also be implemented to other criteria of research such as robotic systems for the purpose of achieving the best performance.

## AUTHORS:

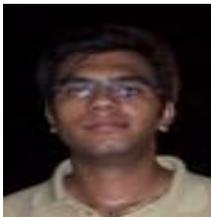
**Kumar Nikhil**, is a final year student IT engineering. Earlier he has done his Diploma in Mechanical engg from Aligarh Muslim University, Aligarh. His areas of interest includes Computer Networks, Software Testing.

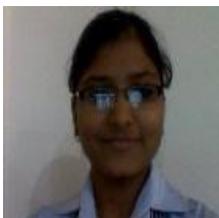
**Swati Agarwal**, topper of her department, is a student of the final year IT engineering and has also published a research paper in a national conference earlier on Neuro Linguistic Programming. Her areas of interest include DBMS, Software Testing and Computer Networks.